\documentclass[preprint,showpacs,preprintnumbers,amsmath,amssymb]{revtex4}
\usepackage{graphicx}
\usepackage{dcolumn}
\usepackage{bm}
\usepackage{mathrsfs}
\usepackage{amsmath,amssymb,graphics}

\begin{document}

\title{Area Spectrum of the Large AdS Black Hole from Quasinormal Modes}

\author{Shao-Wen Wei\footnote{E-mail: weishaow06@lzu.cn}
        and
        Yu-Xiao Liu\footnote{Corresponding author. E-mail:iuyx@lzu.edu.cn}
        }
\affiliation{
    Institute of Theoretical Physics, Lanzhou University,
           Lanzhou 730000, P. R. China}

\begin{abstract}
Using the new physical interpretation of quasinormal modes
proposed by Maggiore, we calculate the area and entropy spectra
for the 3-dimensioal and 5-dimensional large AdS black holes. The
spectra are obtained by imposing the Bohr-Sommerfeld quantization
condition to the adiabatic invariant quantity. With this
semiclassical method, we find that the spacings of the area and
entropy spectra are equidistant and independent of the AdS radius
of the black hole for both the cases. However, the spacings of the
spectra are not the same for different dimension of space-time.
The equidistant area spectra will be broken when the black hole
has other parameters (i.e., charge and angular momentum) or in a
non-Einstein's gravity theory.
\end{abstract}


\maketitle

\section{Introduction}
\label{secIntroduction}

In 1974, Bekenstein presented the famous conjecture
\cite{Bekenstein1}, namely, in a quantum theory, the black hole
area should be represented by a quantum operator with a discrete
spectrum of eigenvalues. Since then, the quantization of the black
hole horizon area has become a more interesting investigation.
Proving that the black hole horizon area is an adiabatic
invariant, Bekenstein obtained an equidistant area spectrum
$\mathcal{A}_{n}=\epsilon \hbar n (n=0,1,2,...)$. While the
spacing $\epsilon$ has been somewhat controversial.


In this direction, an important step was made by Hod about ten
years ago. He suggested that the spacing of the area spectrum
$\epsilon$ can be determined by utilizing the quasi-normal mode
frequencies of an oscillating black hole \cite{Hodprl1998}.
Kunstatter pointed that, for a system with energy $E$ and
vibrational frequency $\Delta\omega(E)$, the ratio
$\frac{E}{\Delta\omega(E)}$ is an adiabatic invariant
\cite{Kunstatterprl2003}. Substituting $M$ for $E$ and identifying
$\Delta \omega(E)$ as the most appropriate choice for the
frequency, the area spectrum $\mathcal{A}_{n}=4 n \ell_{p}^{2} \ln
3$ was obtained by way of the Bohr-sommerfeld quantization
condition. This rejuvenated the interest of investigation for the
quantization of black hole area
\cite{Hodprl1998,Dreyerprl2003,Polychronakosprd2004,Setarecqg2004,
Setareprd2004,Lepeplb2005,Vagenasjhep2008,Medvedcqg2008,
Kothawalaprd2008,Tanaka2008,Ansarinpb2008,Freidelcqg2008,Liplb2009,Cameliaplb2009,0910.5779}.
Recently, Maggiore argued that, in high damping limit, the proper
frequency of the equivalent harmonic oscillator, which is
interpreted as the quasi-normal mode frequencies $\omega(E)$,
should be of the form
$\omega(E)=\sqrt{|\omega_{R}|^{2}+|\omega_{I}|^{2}}$ rather than
the real part $\omega_{R}$ \cite{Maggioreprl2008}. It is easy to
see that when $\omega_{I} \ll \omega_{R}$, one could get
$\omega(E)=\omega_{R}$ approximately which was adopted
extensively, in \cite{Hodprl1998,Setarecqg2004,Dreyerprl2003}.
However, at high excited quasi-normal modes (i.e., $\omega_{R}\ll
\omega_{I}$), thus, one should  get $\omega(E)= |\omega_{I}|$.

Motivated by this idea, the area spectrum was obtained by Medved
and Vagenas \cite{Medvedcqg2008,Vagenasjhep2008}, where they took
the choice that the vibrational frequency
$\Delta\omega(E)=|\omega_{I}|_{m}-|\omega_{I}|_{m-1}$. They found
the area spectrum is non-equidistant and depends on the angular
momentum $J$ of Kerr black hole. It can also be seen that, when
the angular momentum $J\rightarrow0$, the area spectrum is
equidistant. Especially, for the 5-dimensional Gauss-Bonnet(GB)
black hole, Kothawala et.al. first pointed out that the area
spectrum is non-equidistant, but the entropy spectrum is
equidistant \cite{Kothawalaprd2008}. However, when one setting the
Gauss-Bonnet coupling constant $\alpha_{GB}\rightarrow 0$, the
area spectrum would be equidistant. The reason of this may be that
the relationship $S = \frac{\mathcal{A}}{4\hbar}$ between the
horizon area and associated entropy does not hold anymore. In
\cite{Weijhep2008}, we conjecture that, for the non-rotating black
holes with no charge, the entropy spectrum is equidistant and is
independent of the dimension of space-time. However, the spacing
of the area spectrum depends on gravity theory. In Einstein's
gravity, it is equally spaced, otherwise it is non-equidistant.
For a charging or rotating black hole (e.g., the Kerr black hole),
the spacing of area spectrum will not be equal because of the
existence of the angular momentum $J$ or the charge $Q$.

In this paper, we will calculate the area and entropy spectra of
large  Ads black hole using the asymptotic quasinormal modes of
obtained from \cite{Musiriplb2003}. The quasinormal modes for
Schwarzschild-Ads black hole were also obtained in
\cite{Natario2004,Daghighcqg2009}, the numerical result can be found
in \cite{Cardosoprd2003}. The Stability of simply rotating
Myers-Perry-AdS black holes and higher dimensional R-N-Ads black
hole were discussed in \cite{Konoplyaprd2008}. Especially, in
\cite{Daghighcqg2009}, the highly real quasinormal modes has a
logarithm term in the imaginary part rather than the real part,
which is different from the Schwarzschild case. However, it seems
has no affect on obtaining a equidistant spectra.

The paper is organized as follows. In Section \ref{AdS black
hole}, we briefly review the thermodynamics and the quasinormal
modes of the large AdS black hole. In Section \ref{3d area
spectrum}, we will apply the semiclassical method to the
3-dimensional lager AdS black hole and obtain an equidistant area
spectrum. The area spectrum of the 5-dimensional lager AdS black
hole is obtained in Section \ref{5darea spectrum}, which is also
equally spaced. Finally, the paper ends with a brief Summary.

\section{Quasinormal modes of the large AdS black hole}
\label{AdS black hole}

In this section, we give a brief review of the AdS black hole. For a
$d$-dimensional AdS black hole, the metric can be written as
\begin{equation}
  ds^2 =-f(r) dt^2 +\frac{1}{f(r)}dr^2 +r^2 d\Omega_{d-2}^2.
\end{equation}
The metric function is
\begin{equation}
 f(r) = 1+\frac{r^2}{l^2}-\frac{\varpi_{d-1} M}{r^{d-3}},
\end{equation}
where $l$ and $M$ are the AdS radius and mass of the black hole,
respectively. For a large black hole, the metric function $f(r)$ is
simplified to
\begin{equation}
 f(r) =\frac{r^2}{l^2}
        -\frac{\varpi_{d-1} M}{r^{d-3}}.
\end{equation}
The event horizon is located at $f(r_{h})=0$ and the radius $r_{h}$
satisfies
\begin{equation}
 r_{h}=(l^{2}\varpi_{d-1} M)^{\frac{1}{d-1}}.
\end{equation}
The Hawking temperature is
\begin{equation}
 T=\frac{d-1}{4\pi}\frac{r_{h}}{l^{2}}.
\end{equation}
The first law of black hole thermodynamics is
\begin{equation}
 dM=TdS.
\end{equation}
Performing the partial wave decomposition $\Phi=e^{i(\omega
t-\vec{p}\cdot \vec{x})}\Psi(r)$ and solving the massless scalar
wave equation with proper boundary conditions, the quasi-normal
modes can be obtained \cite{Musiriplb2003}. For the $d=3$ case,
the quasi-normal modes are read as
\begin{equation}
 \frac{\omega}{4\pi T}=\pm \hat{p}-i m,\;\;m=1,2,..., \label{quasinormal1}
\end{equation}
where $\hat{p}$ is a dimensionless variable and given by
\begin{equation}
\hat{p}^{2}=\frac{p}{4\pi l T}.
\end{equation}
For the case $d=5$, the quasi-normal modes are
\begin{equation}
 \frac{\omega}{\pi T}=2m (\pm 1-i),\;\;m=1,2,....\label{quasinormal2}
\end{equation}
It is easy to see that, for both the cases, the quasi-normal modes
$\omega$ are complex frequencies. In fact, this is a universal
character of quasi-normal modes. For the quasi-normal modes
$\omega$ appeared in the wave function as $e^{i\omega t}$, the
image part will decrease to zero at infinite. The quasi-normal
modes are also generally believed to depend only on the black hole
parameters (i.e., the mass $M$, charge $Q$ and angular momentum
$J$). It can be proved that the quasi-normal modes
(\ref{quasinormal1}) and (\ref{quasinormal2}) only depend on the
mass $M$ of the large AdS black hole.

\section{Area spectrum of the 3-dimensional large AdS black hole}
\label{3d area spectrum}

In this section, we first give a general statement of the method we
used. Then we apply it to the 3-dimensional large AdS black hole and
obtain the area and entropy spectra, respectively.

Several years ago, Kunstatter proposed that given a system with
energy $E$ and vibrational frequency $\Delta \omega(E)$, a nature
adiabatic invariant quantity is \cite{Kunstatterprl2003}:
\begin{eqnarray}
 I=\int \frac{dE}{\Delta \omega(E)}. \label{condition}
\end{eqnarray}
At the large $n$ limit, the Bohr-Sommerfeld quantization can be
expressed as
\begin{eqnarray}
 I\approx n\hbar.
\end{eqnarray}
One may argue, the quantum number $n$ should replace by
$(n+\frac{1}{2})$ like harmonic oscillator. Then, one could discuss
the remnant of black hole. However, we need keep in mind that the
number $n$ is a large number and $n\sim (n+\frac{1}{2})$. So, it is
not proper to discuss the remnant of black hole under this case.
Adiabatic invariant quantity $I$ is
\begin{equation}
 I=\int \frac{dE}{\Delta\omega}=\int \frac{dM}{\Delta\omega}, \label{BTZ}
\end{equation}
where we have identified the energy $E$ of this black hole system
as it's mass $M$ in the second step. Through the interpretation
that identifying the vibrational frequency $\Delta\omega(E)$ as
the quasinormal modes, much work has been done (e.g.,
\cite{Hodprl1998,Setarecqg2004,Dreyerprl2003}), where the
vibrational frequency $\Delta\omega(E)$ was regarded as the real
part of the quasinormal modes. Recently, Maggiore refined Hod's
treatment by arguing that the physically relevant frequency would
actually be \cite{Maggioreprl2008}
\begin{equation}
 \omega(E)=\sqrt{|\omega_{R}|^{2}+|\omega_{I}|^{2}},
\end{equation}
where $\omega_{R}$ and $\omega_{I}$ are the real and imaginary
parts of the quasinormal modes frequency respectively. When
$\omega_{I}\rightarrow 0$, one could get $\omega(E)=|\omega_{R}|$
approximately. However, at the case of large $m$ or highly excited
quasinormal modes for which $\omega_{R}\ll \omega_{I}$, the
frequency of the harmonic oscillator becomes $\omega(E)=
|\omega_{I}|$. Under this supposition, work has been applied to
different black holes. The results show that, for the non-rotating
black holes with no charge, the area and entropy spectra are both
equally spaced in Einstein's gravity \cite{Weijhep2008}. For other
gravity theory, the result is partially hold. Take the Gauss
Bonnet gravity as a example, it is first pointed in
\cite{Kothawalaprd2008} that the area spectrum is not equally
spaced, but the entropy is equidistant still. If the black hole
has other parameters, i.e., the angular momentum $J$, the
equidistant area spectrum will be broken
\cite{Vagenasjhep2008,Medvedcqg2008}. However, if setting the
angular momentum $J\rightarrow 0$, the area spectrum will tend to
an equally spaced spectrum.

For the case $d$=3, the radius of event horizon, the area and the
Hawking temperature are given, respectively
\begin{eqnarray}
 r_{h}&=& l (\varpi_{2} M)^{\frac{1}{2}},\\
 \mathcal{A}    &=&2\pi r_{h}, \label{areaform1}\\
 T    &=&\frac{r_{h}}{2\pi l^{2}}.
\end{eqnarray}
At the large $m$ limit, the vibrational frequency $\Delta\omega$
is
\begin{equation}
 \Delta\omega=|\omega_{I}|_{m}-|\omega_{I}|_{m-1}=\frac{2r_{h}}{l^{2}}.
\end{equation}
Substituting $\Delta\omega$ into (\ref{BTZ}), we obtain the
adiabatic invariant quantity
\begin{eqnarray}
 I&=&\int \frac{dM}{\Delta\omega}\nonumber\\
   &=&\int \frac{1}{\frac{dr_{h}}{dM}} \frac{dr_{h}}{\Delta\omega}\nonumber\\
   &=& \frac{r_{h}}{\varpi_{2}}.
\end{eqnarray}
Using the Bohr-Sommerfeld quantization condition
(\ref{condition}), at the large $n$ limit, we obtain
\begin{equation}
\frac{r_{h}}{\varpi_{2}}=n \hbar.
\end{equation}
Recalling the area from (\ref{areaform1}), the area spectrum of this
black hole is given
\begin{equation}
 \mathcal{A}_{n}=2\pi \hbar \varpi_{2}\cdot n . \label{BTZresult}
\end{equation}
It is clear that this area spectrum is equally spaced with
equidistant $\Delta \mathcal{A}=2\pi \hbar \varpi_{2}$. The area
spectrum is independent of the AdS radius $l$ of the black hole.
Recalling the relationship $S=\frac{\mathcal{A}}{4 \hbar}$ between
horizon area and associated entropy, one could get the entropy
spectrum
\begin{equation}
 S_{n}=\frac{\pi \varpi_{2}}{2} \cdot n,
\end{equation}
with the spacing
\begin{equation}
 \Delta S=S_{n+1}-S_{n}=\frac{\pi \varpi_{2}}{2}.
\end{equation}
The entropy is also equidistant and the spacing is independent of
the AdS radius $l$ of the black hole.

\section{Area spectrum of the 5-dimensional large AdS black hole}
\label{5darea spectrum}

In this section, we would like to focus on the 5-dimensional large
AdS black hole and obtain the area and entropy spectra.

For the case $d$=5, the radius of event horizon, the area and the
Hawking temperature are given, respectively
\begin{eqnarray}
 r_{h}&=& (l^{2}\varpi_{d-1} M)^{\frac{1}{4}},\\
 \mathcal{A}    &=&2\pi^{2} r_{h}^{3}, \label{areaform2}\\
 T    &=&\frac{r_{h}}{\pi l^{2}}.
\end{eqnarray}
For a general $m$, the vibrational frequency $\Delta\omega$ can be
obtained
\begin{eqnarray}
 \Delta\omega &=&\sqrt{(|\omega_{R}|_m)^{2}+(|\omega_{I}|_m)^{2}}
               \nonumber \\
              &-&\sqrt{(|\omega_{R}|_{m-1})^{2}+(|\omega_{I}|_{m-1})^{2}}
              =\frac{2\sqrt{2}r_{h}}{l^{2}}.
\end{eqnarray}
Substituting $\Delta\omega$ into (\ref{BTZ}), we obtain the
adiabatic invariant quantity
\begin{eqnarray}
 I&=&\int \frac{dM}{\Delta\omega}\nonumber\\
   &=&\int \frac{1}{\frac{dr_{h}}{dM}} \frac{dr_{h}}{\Delta\omega}\nonumber\\
   &=& \frac{\sqrt{2}r_{h}^{3}}{3\varpi_{4}}.
\end{eqnarray}
Using the Bohr-Sommerfeld quantization condition
(\ref{condition}), at the large $n$ limit, we obtain
\begin{equation}
\frac{\sqrt{2}r_{h}^{3}}{3\varpi_{4}}=n \hbar.
\end{equation}
Recalling the area from (\ref{areaform2}), the area spectrum are
\begin{equation}
 \mathcal{A}_{n}=3\sqrt{2}\pi^{2} \varpi_{4} \hbar\cdot n .
\end{equation}
This area spectrum is also equally spaced and with equidistant
spacing $\Delta \mathcal{A}=3\sqrt{2}\pi^{2} \varpi_{4} \hbar$.
For this case, it shares the same character with the case $d$=3
that the area spectrum is independent of the AdS radius $l$ of the
black hole. However, the spacings of the area spectra are not
equal, which is due to the different dimension of space-time. Then
the entropy spectrum can also be obtained
\begin{equation}
 S_{n}=\frac{3\sqrt{2}\pi^{2} \varpi_{4}}{4} \cdot n,
\end{equation}
with the spacing
\begin{equation}
 \Delta S=S_{n+1}-S_{n}=\frac{3\sqrt{2}\pi^{2} \varpi_{4}}{4}.
\end{equation}
The entropy is also equidistant and the spacing is independent of
the AdS radius $l$ of the black hole.

In fact, the spacing of area and entropy spectra for d-dimensional
Schwarzschild-Ads black hole are also obtained in
\cite{Daghighcqg2009}
\begin{eqnarray}
 &&\Delta A=16\pi \sin(\frac{\pi}{D-1}),\nonumber\\
 &&\Delta S=4\pi \sin(\frac{\pi}{D-1}).
\end{eqnarray}
If we choose the proper value of $\varpi$, i.e. $\varpi_{2}=8$ and
$\varpi_{4}=\frac{4}{3\pi}$, it will reduce to our cases. Using the
asymptotic quasinormal modes given in \cite{Natario2004}, one could
see that the result is the same.

\begin{table*}
\begin{tabular}{|c|c|c|}
  \hline
   & $\Delta{\mathcal{A}}_n$ & $\Delta{\mathcal{S}}_n$  \\
  \hline
   3D BTZ black hole \cite{Weijhep2008}  & $\surd$ & $\surd$ \\
  \hline
   3D Spinning black hole(neglecting the angular moment)  \cite{Fernando2009}        & $\surd$ & $\surd$ \\
  \hline
   4D Schwarzchild  black hole    & $\surd$ & $\surd$ \\
  \hline
   4D Schwarzschild-AdS black hole \cite{Liplb2009} & $\surd$ & $\surd$ \\
  \hline
   3D and 5D large AdS black hole & $\surd$ & $\surd$ \\
  \hline
   4D Kerr black hole  \cite{Vagenasjhep2008,Medvedcqg2008}  & $\times$ & $\times$ \\
  \hline
   5D GB black hole \cite{Kothawalaprd2008}    & $\times$ & $\surd$ \\
  \hline
\end{tabular}
\label{table}
  \caption{ The spacings of the area spectra ${\mathcal{A}}_n$ and entropy spectra ${\mathcal{S}}_n$
             for virous black holes. $\surd$ and $\times$ stand for that the
             spectrum is equidistant and non-equidistant, respectively.
  }
\end{table*}

\section{Summary and Outlook}
\label{Summary}

In this paper, we succeed in utilizing a new physical
interpretation of quasinormal modes to the large Ads black hole
following the Kunstatter¡¯s method. By modifying the frequency
$\omega(E)$ appeared in the adiabatic invariant of black hole and
using the Bohr-Sommerfeld quantization at the large $n$ limit, we
investigate the area and entropy spectra of 3-dimensional and
5-dimensional large AdS black holes. For both the cases, the
spectra are equidistant and independent of the AdS radius $l$.
However, the spacings of the spectra are different, which is
because of the different dimension of space-time. Although our
results are still somewhat speculative, they certainly propose a
reasonable physical interpretation on the spectrum of the large
AdS black hole quasinormal modes. Our results still suppose the
conjecture that, for the non-rotating black holes with no charge,
the spacings of the area and entropy spectra are equidistant and
are independent of the dimension of space-time. From the Table I,
it easy to see that the extra parameter (angular momentum $J$) of
black hole will broken the equidistant spectra. So, it is our
conjecture that, for a Reissner-Norstr\"om (or Kerr-Newmann) black
hole, the area spectrum will not equidistant. But if setting the
charge $Q$ (and angular momentum $J$) $\rightarrow 0$, the
equidistant area and entropy spectra will reoccur. This needs to
be proved in future.

\section*{Acknowledgements}

This work was supported by Program for New Century Excellent
Talents in University, the National Natural Science Foundation of
China(NSFC)(No. 10705013), the Doctoral Program Foundation of
Institutions of Higher Education of China (No. 20070730055), the
Key Project of Chinese Ministry of Education (No. 109153) and the
Fundamental Research Fund for Physics and Mathematics of Lanzhou
University (No. Lzu07002).

\end{document}